# Generating Keys in Elliptic Curve Cryptosystems

**Dragan Vidakovic and Dusko Parezanovic**
Gimnazija, Ivanjica, Serbia

## ABSTRACT

In this paper, we will present how to find keys elliptic curve cryptosystems (ECC) with simple tools of Delphi **7** console application, using the software problem solving of the scalar point multiplication in the field GF (p), where p is an arbitrary prime number.

## 1. INTRODUCTION

Since our permanent goal is to have more of our own software [1], we have decided to present in this paper the software implementation of the scalar point multiplication in a finite field $F_p$ [2]. Besides this basic motive, there is one more which seems to be more important – the actual prioritizing ECC over RSA. If we compare ECC to RSA, we can conclude that ECC is faster, safer and requires significantly less resources (ranging from energetic resources, up to hardware). So, ECC becomes a better alternative when it comes to authentication of different mobile devices, in sensor networks, and wherever there is a need (for a variety of reasons) to avoid the “raw” and endless increasing of the number of bits, as it is the case in RSA [3], [4].

There is one more reason why we have opted for this work: despite the fact that mathematical bases of ECC and RSA are completely different, they are solved using the same instruments, i.e. tools we have developed [5]. The set of tools necessary for the formation of a private key in ECC is a real subset of tools needed for the solving of the same problem in RSA.

Of course, we cannot ignore the fact ( in our intention to introduce a scalar point multiplication) that both leading schemes of digital signature (ECC and RSA) are brilliant examples of the application of the theory of finite fields [6 - 9], where ECC [10-17] is still based on the richness of algebraic features of elliptic curves , while RSA is based on the assumption of difficult factoring of large numbers whose only factors are two large (probably) prime numbers.

## 2. TASK AND AIM

In this paper we do not intend to deal with the theory of elliptic curves, because there is a lot of good work on that [2],[8], [18-19]. We are interested in the results of the theory that we wish to encode.

Our goal is to calculate k*P=Q, where P and Q are points (set E) on an eliptic curve: $y^2= x^3 +ax +b$. Operation '*' denotes the series of Point doubling and Point adding.

### 2.1 Point adding

For given two points $P(x_p,y_p)$, $Q(x_q,y_q)$ $(P\neq\pm Q)$ in the set E, the group operator will allow us to calculate a third point $R(x_r,y_r)$, also in the set E, such that P + Q = R.

Not difficult to find the coordinates of point R: $x_r = s^2 - x_p - x_q$ where $s^2 = 2x_p + x_q + x_r - x_p$

As point R belongs to the straight line (PQ) then $s= \frac{y_r - y_p}{x_r - x_p}$ and we find: $y_r = y_p + s(x_r - x_p)$

### 2.2 Point doubling

For given point P(xp,yp) in the set E, the group operator will allow us to calculate a third point $R(x_r,y_r)$, also in the set E, such that P + P = 2P= R.

$x_r = s2 - 2x_p$ where $s= \frac{3x_p^2 - a}{2y_p}$ and $y_r= y_p + s(x_r - x_p)$

### 2.3 Point multiplication

One of the most important operations for all applications of elliptic curves is scalar multiplication. Scalar multiplication consists of computing the value of a large integer multiplied by a point by doing a series of point doublings and additions until the product point is reached. In this paper we will use approach for computing k*P was introduced by Montgomery [20].

**Algorithm: Binary method**

INPUT: An integer k>0 and a point P.
OUTPUT: Q=k*P
1. Set $k \leftarrow (k_{l-1} \ldots k_1 k_0)_2$
2. Set $P_1 \leftarrow P$, $P_2=2P$.
3. for I from l-2 downto 0 do
   If $k_i=1$ then
      Set $P_1 \leftarrow P_1 + P_2$, $P_2 \leftarrow 2P$.
   Else
      Set $P_2 \leftarrow P_2 + P_1$, $P_1 \leftarrow 2P_1$.

4. RETURN ($Q=P_1$)

## 3. CONSOLE APPLICATION DELPHI 7

In this paragraph, we will encode the algorithm above by a simple tool, and that is the Console Application Delfi 7 (see Appendix). Our goal is not to develop the program that is characterized by faultless performance; our goal is to show that this program does its task (point multiplication) and that it is possible to solve such a serious task in a simple way which is good for learning and entering the problem, the way that breaks the concern about the cryptic of cryptography itself. This way can assure us that applied cryptography is not as difficult as it seems to be. This program is a really good base for those who want to optimize and develop their own tool that won't be just illustrative, but it will be able to serve, considering the fact that we work with arbitrarily large numbers for which it is only necessary to set the initial constant in the Unit [21-22].

### 3.1 Example (Outputs)

NIST (National Institute of Standards and Technology) recommended ten finite fields [2]. We will observe the prime finite field $F_p$ for $p=2^{192} - 2^{64} -1$

For curve P-192 a= - 3 and base point:

$P(X_G, Y_G)$:
$X_G$ = 0x 188da80eb03090f67cbf20eb43a118800f4ff0afd82ff1012
$Y_G$= 0x 07192b95ffc8da78631011ed6b24cdd53f977a11e794811
Let k=
1000000000000000000000000000000000000000000000000000000000010
0000000000000000000000000000000000000000000000001000000000000
00000000000000000000000000010001101011

(probably) prime large (160 bit) number. ( k[159]=1, k[100]=1, k[50]=1,k[10]:=1; k[6]:=1; k[5]:=1; k[3]=1, k[1]=1, k[0]=1). This number is not random "good" for cryptography, but was taken to illustrate the example.

Coordinates of the point $Q(X, Y) = k * P(X_G, Y_G)$ are :

X: 0x a9355c37074c8195faa23d1d071997fe4ea7a2bdec781047

Y:0x 911a232aabeba33b5e8f29743f2837955cd5bf1f74aa9a24

## 4. ECDSA KEY PAIR

The Elliptic Curve Digital Signature Algorithm (ECDSA) [2] is the elliptic curve analogue of DSA (Digital Signature Algorithm) [22].

Security of DSA based on the computational intractability of the discrete logarithm problem (DLP) [23].

The mathematical basis for the security of elliptic curve cryptosystems (ECC) is the computational intractability of the elliptic curve discrete logarithm problem (ECDLP) [24-25].
To put it simply: It is dificult to find a point P and integer k, given their product k*P.

### 4.1 ECC parameters

**1. Public key is Q.**

**2.Private key is k:** (we choose)

1000000000000000000000000000000000000000000000000000000000010
0000000000000000000000000000000000000000000000001000000000000
00000000000000000000000000010001101011

**3. Modulus:**

1111111111111111111111111111111111111111111111111111111111111
1111111111111111111111111111111111111111111111111111111111111
1111101111111111111111111111111111111111111111111111111111111
111111111)
Which we generate the key pair for ECDSA.

### 4.2 RSA parameters - comparison

Elliptic curve cryptography can provide the same level and type security as RSA but with much shorter keys. More precisely, ECC takes one-sixth computational effort to provide the same security that ones get with 1024-bit RSA, what no comments confirmed the following example:

**1024 -bit RSA parameters**

**1. Public key pair (n,e)**

**e**:111 (we choose)
**n**: (we count)
1000000000000000000000000000000000000000000000000000000000000
0000000000000000000000000000000000000000000000000000000000000
0000000000000000000000000000000000000000000000000000000000000
0000000000000000000000000000000000000000000000000000000000000
0000000000010000000000000000000000000000000000000000000000000
0000000000000000000000000000000000000000000000000000000000000
0000000000000000010000000000000000000000000010000000000000000
0000000000000000000000000000000000100000000000000000000000000
0000000000001101001000010000000000000000000000000000000000000
0000000000000000000000000000000000000000000000000000000000000
0000000000000000000000000000100000000000000000000000000100000
0000000000000000000000000000000000000000000001000000000000000
0000000000000000000000011010010000010000000000000000000000000

1000000000000000000000000100000000000000000001000000000
0000000000000000100000000011010010000000000000000011010
0100000000000000000000000000000000000000111100000110000
0000000000000000000001010100010101001010111

**2. Private key d:** (we count)

1001001001001001001001001001001001001001001001001001001
0010010010010010010010010010010010010010010010010010010
0100100100100100100100100100100100100100100100100100100
1001001001001001001001001001001001001001001001001001001
0010010010100100100100100100100100100100100100100100100
1001001001001001001001001001001001001001001001001001001
0010010010010010100100100100100100100100110110110110110
1101101101101101101101101101101110010010010010010010010
0100100100110011100100001001001001001001001001001001001
0010010010010010010010010010010010010010010010010010010
0100100100100100100100100100110110110110110110111001001
0010010010010010010010010010010010010010010110110110110
1101101101101101101101111100101101010010010010010010010
1001001001001001001001001001011011011011011011100000000000
0000000000000001001001001011000001000100100100100110011
1001000000000000000000000000000000000000100010010100100
1001001001001001001001111001010001011111

**3. Modulus: (**we count)

1000000000000000000000000000000000000000000000000000000
0000000000000000000000000000000000000000000000000000000
0000000000000000000000000000000000000000000000000000000
0000000000000000000000000000000000000000000000000000000
0000000000100000000000000000000000000000000000000000000
0000000000000000000000000000000000000000000000000000000
0000000000000001000000000000000000000000100000000000000
0000000000000000000000000000000100000000000000000000000
0000000000110100011110100000000000000000000000000000000
0000000000000000000000000000000000000000000000000000000
0000000000000000000000000010000000000000000000000010000
0000000000000000000000000000000000000000010000000000000
0000000000000000000001101000111101000000000000000000000
0100000000000000000000000010000000000000000000100000000
0000000000000000100000000001101000111000000000000001101
0001111000000000000000000000000000000000001111000010000
00000000000000000000001010100001110100111000.)

## 5. CONCLUSION

Our paper, as well as other ones, is on the edge of the idea that the best protection is by using our own tools. In order to achieve this, it is necessary to increase interest in cryptography [26], and, in our opinion, the best way to do that is to encode and test cryptographic algorithms.

That is the main reason why we have opted for the simplest console application. We want to show that it is possible to solve such a significant task (the scalar point multiplication in a finite field Fp) in an almost elementary way, without any software-hardware tools.

Thus, the problem becomes more comprehensible to a wider range of readers, and in a simple way it demonstrates that the applied cryptography is not as unavailable as it seems to be, or as it is presented.

The false idea of mysterious and too complicated cryptography reduces interest, refuses and discourages young researchers at the very beginning of a work, and thus it significantly minimizes the chances for developing our own tools. Therefore, we become dependent on the software producers, for whom our secrets become absolutely available. And each country that intends to protect itself as much as possible must take into account that fact, so as to protect itself from those who protect our secrets from all the others, except from them themselves.

## APPENDIX

### A. NIST recommended prime finite field $F_p$ for $p=2^{192} - 2^{64} - 1$

```
program multiplikacija_nist;
{$APPTYPE CONSOLE}
uses
 SysUtils,
{ Unit2mult_200 in '..\..\..\Bin\Unit2mult_200.pas', }
 multiplikacija_dalje in 'multiplikacija_dalje.pas';
label 1;
var pxp,pyp,p1x,p1y,p2x,p2y,x2,y2,rez1,rez2:array [0..nn] of integer;
```

```
 var modu,a,b,k:array [0..nn] of integer;
 i,j,s1,l,i1,s,p:integer;
begin
b[64]:=1;
 a[0]:=1;
 modu[192]:=1; oduzmi(modu,a,modu);
{procedure oduzmi- subtraction}
 oduzmi(modu,b,modu);
 a[1]:=1;
{XG – coordinates}
 pxp[12]:=1; pxp[16]:=1; pxp[17]:=1; pxp[18]:=1;
 pxp[19]:=1; pxp[20]:=1; pxp[21]:=1; pxp[22]:=1;
 pxp[23]:=1; pxp[25]:=1; pxp[31]:=1; pxp[32]:=1;
 pxp[34]:=1; pxp[35]:=1; pxp[36]:=1; pxp[37]:=1;
 pxp[38]:=1; pxp[39]:=1; pxp[41]:=1; pxp[43]:=1;
 pxp[48]:=1; pxp[49]:=1; pxp[50]:=1;pxp[51]:=1;
 pxp[52]:=1; pxp[53]:=1; pxp[54]:=1; pxp[55]:=1;
 pxp[58]:=1; pxp[60]:=1; pxp[61]:=1; pxp[62]:=1;
 pxp[63]:=1;
{YG – coordinates}
 pyp[15]:=1; pyp[17]:=1; pyp[20]:=1; pyp[21]:=1;
 pyp[22]:=1; pyp[23]:=1; pyp[26]:=1; pyp[27]:=1;
 pyp[28]:=1; pyp[29]:=1; pyp[33]:=1; pyp[38]:=1;
 pyp[40]:=1; pyp[41]:=1; pyp[42]:=1; pyp[43]:=1;
 pyp[45]:=1; pyp[46]:=1; pyp[47]:=1; pyp[49]:=1;
 pyp[52]:=1; pyp[53]:=1; pyp[54]:=1; pyp[55]:=1;
 pyp[56]:=1; pyp[57]:=1; pyp[58]:=1; pyp[61]:=1;
 pyp[63]:=1; pyp[65]:=1; pyp[67]:=1;
{(probably) prime number k}
k[159]:=1,k[100]:=1;k[50]:=1;k[10]:=1;k[6]:=1;k[5]:=1;k[3]:=1;k[1]:=1;
k[0]:=1;
 s1:=0;
 for i:=0 to nn do
 begin
 p1x[i]:=pxp[i];
 p1y[i]:=pyp[i];
 end;
{Point multiplication}
duplope(pxp,pyp,a,modu,x2,y2);
 for i:=0 to nn do
 begin
 p2x[i]:=x2[i];
 p2y[i]:=y2[i];
 end;
```

```
dokle(k,s1);
for i:=s1-1 downto 0 do
begin
 if k[i]=1 then
 begin
 s:=1;
 koji(p1x,p2x,s);
 if (s=0) then
 begin
{procedure duplope- point doubling}
 duplope(p2x,p2y,a,modu,p2x,p2y);
 goto 1;
 exit;
 end;
{procedure dverazne-point addition}
 dverazne(p1x,p1y,p2x,p2y,modu,p1x,p1y);
 duplope(p2x,p2y,a,modu,p2x,p2y);
 end
 else
 begin
 s:=1;
 koji(p1x,p2x,s);
 if (s=0) then
begin
 duplope(p1x,p1y,a,modu,p1x,p1y);
 goto 1;
 exit;
 end;
 dverazne(p2x,p2y,p1x,p1y,modu,p2x,p2y);
 duplope(p1x,p1y,a,modu,p1x,p1y);
 end;
 end;
{procedure pisi- bin to hex}
 pisi(p1x);
 writeln;
 pisi(p1y);
 readln;
 1:  begin
 if s=0 then
 write(' Infiniti Point ');
 end; end.
```

**Remark**: Appropriate procedures are in [27]